\documentstyle[12pt]{article}

\newcommand{\be}{\begin{equation}}
\newcommand{\ee}{\end{equation}}

\newcommand{\ba}{\begin{eqnarray}}
\newcommand{\ea}{\end{eqnarray}}

\newcommand{\la}{\lambda}
\newcommand{\al}{\alpha}

\newcommand{\r}{\rho}

\newcommand{\Tr}{\rm Tr}
\newcommand{\tr}{\rm Tr}

\begin{document}

\hsize36truepc\vsize51truepc
\hoffset=-.4truein\voffset=-0.5truein
\setlength{\textheight}{8.5 in}

\begin{titlepage}
\begin{center}
\vskip 0.6 in
{\large  Characteristic polynomials of random matrices
at edge singularities}
\vskip .6 in
       {\bf Edouard Br\'ezin \footnote{$  ${\it
 Laboratoire de Physique Th\'eorique de l'\'Ecole Normale
Sup\'erieure, Unit\'e Mixte de Recherche 8549 du Centre National de la
Recherche
Scientifique et de l'\'Ecole Normale Sup\'erieure,
24 rue Lhomond, 75231 Paris Cedex 05, France.\\ {\bf
brezin@physique.ens.fr}}}}
            {\it{and}}\hskip 0.3cm
       {\bf {Shinobu Hikami}}  \footnote{$  ${\it{ Department of
Basic Sciences, University of Tokyo, Meguro-ku, Komaba 3-8-1, Tokyo
153,
Japan.\\ {\bf hikami@rishon.c.u-tokyo.ac.jp}}}}\\

 \vskip 0.5 cm
{\bf Abstract}
\end{center}
\vskip 14.5pt

{\leftskip=0.5truein\rightskip=0.5truein\noindent
{\small

We have discussed earlier the correlation functions of the random variables
$\det(\la-X)$ in which $X$ is a random matrix. In particular the moments of
the distribution
of these random variables are universal functions, when measured in the
appropriate units of the level spacing.
When the $\la$'s, instead of belonging to the bulk of the spectrum,
approach the edge, a cross-over takes
place to an Airy or to a Bessel problem, and we consider here these
modified classes of universality.

Furthermore, when an external matrix source is added to the probability
distribution of $X$, various new
phenomenons may occur and one can tune the spectrum of this source matrix
to new critical points. Again there are
remarkably simple formulae for arbitrary source matrices, which allow us to
compute the moments of the characteristic polynomials
in these cases as well.
}
\par}

\end{titlepage}
\setlength{\baselineskip}{1.5\baselineskip}
\section{Introduction}
   In the theory of random matrices , the n-point correlation
   functions of the eigenvalues are known to be expressible  as the
determinant of a
two-point  kernel \cite{Dyson,Mehta}. The expressions for those kernels
depend on
 the various classes of
universality :
it is a simple sine-kernel within  the bulk of unitary invariant ensembles,
an Airy kernel at the edge
  of the spectrum or a Bessel kernel for other invariance properties of the
measure.
   The level spacing probability $p(s)$,  has also been computed recently
for  those
   different kernels \cite{Mehta,TracyWidom}.

Another interesting object is given by the average moments
   of the characteristic
   polynomial of the random matrix. These characteristic polynomials
   have been first investigated in \cite{Keating, ConreyFarmer} for a
uniform probability measure on unitary
matrices, in  connection with the moments of the
   Riemann zeta-function.
    These results have been generalized to  random hermitian $N\times N$
matrices
$X$ with a unitary invariant probability measure
\be
    P(X) = \frac{1}{Z}\exp{-N \Tr V(X)}.
   \ee   Explicit formulae for the $2K$-point functions
\be
     F_{2K} (\la_1,\cdots,\la_{2K}) = <\prod_1^{2K}\det(\la_l - X)>
   \ee
   have been derived, which show that these functions are universal in the
Dyson limit, in which the size
N of the matrices goes to infinity, the distances between the $\la$'s go to
zero, and the products
$N(\la_i-\la_j)$ remain finite. In particular the moments
 \be
     F_{2K} (\la,\cdots,\la) = <[\det(\la - M)]^{2K}>
   \ee
of the distribution of the characteristic polynomials were
 given in the large N limit by \cite{BH1a,BH1b}

   \be \label{moment} \exp-{(NKV(\la))}F_{2K}(\la,\cdots,\la) = (2\pi
N\r(\la))^{K^2}
e^{-NK}\gamma_K , \ee
with
\be\label{gammak} \gamma_K = \prod_0^{K-1}\frac{l!}{(K+l)!}\ ,\ee
provided $\la$ belongs to the bulk of the support of the distribution
of the eigenvalues, i.e. provided $\rho(\la)$ does not vanish. Then
 one sees explicitely that the only dependence upon the probability measure
is  through the average
density of eigenvalues
$\r(\la)$, and even the coefficient  $\gamma_K$ is a universal number.

However the result  does take different forms
 for  different universality classes. Our previous investigations
 for the three classical Lie groups, U(N), Sp(N) and O(N), are extended
 here to the Bessel kernel and Airy kernel, for which the density of states
$\r(\la)$
presents
  a singularity at the edge of the spectrum.
Furthermore we have considered a Gaussian case in which an  external
 matrix source
is present\cite{BH2} in the probability distribution of the matrix
\be
    P(X) = \frac{1}{Z}\exp{(-N \Tr \frac{1}{2} X^2 + N \Tr AX)}.
\ee Explicit and simple formulae will be derived here again for the
correlation functions
and the moments of the characteristic polynomials of the matrix $X$, which
depend on the
spectrum of the matrix $A$. By
tuning the spectrum of $A$ appropriately, one can generate a number of
different
situations. For instance, we have investigated in the past
the case in which the average spectrum of $X$ presents  a gap in the
presence of $A$, and by tuning $A$ one can
study the critical point at which this gap vanishes. This creates again a
new class of universality,
and a new
kernel \cite{BH3,BH4}. Other cases, such as 2D gravity in the  double
scaling limit,
   or the Penner model,  would  certainly be of interest as well.

\section{  Sine-kernel}
  For completeness, and for later use, we  begin with the bulk unitary
case, governed by the
sine-kernel, but with  a derivation which differs from our previous
one \cite{BH1a}.  An interesting geometric interpretation of this
problem will also be provided.
The
kernel, from which all the correlation functions may be obtained, is
given in terms of orthogonal polynomials for finite N, but
reduces in the Dyson large N-limit
to the
  sine-kernel
\be
    K(x,y) = \frac{\sin ( x - y)}{x - y}
  \ee
in which $x$ and $y$ are the eigenvalues measured in the scale  of the average
spacing  $\left(2\pi \rho(\la) N\right)^{-1}$. Then, one obtains the
normalized moments

  \be
     I_K = e^{-NKV(\la)}\  \frac{<[\det(\la - X)]^{2K}>}{
    (2\pi
N\r(\la))^{K^2}} = \lim_{\la_i \rightarrow 0} \frac{\det K
(\la_i,\la_j)}{\Delta^2(\Lambda
)}
  \ee
  where $\Delta(\Lambda) = \prod_{i<j} ( \la_i - \la_j)$ and $i,j =
  1,...,K.$
  The r.h.s. may be  expressed  as a contour integral, following eq.(52) of
\cite{BH1a},
  \be
  I_K = \oint\oint \prod_i^K\frac{du_idv_i}{(2\pi i)^2}
\frac{\Delta(U)\Delta(V)
}{\prod_{i=1}^K u_i^K \prod_{i=1}^K
  v_i^K} \prod_{i=1}^K \frac{\sin ( u_i - v_i)}{u_i - v_i}
  \ee
  This may be  further reduced to
  \be
  I_K = \det ( a_{nm})
  \ee
 \be\label{sinedet}
     a_{nm} = \frac{1}{n! m!} \frac{\partial^n}{\partial u^n}
     \frac{\partial^m}{\partial v^m} \frac{\sin(u - v)}{u - v}|_{u=v=0}
   \ee
   where $n,m = 0, 1, ..., K-1$.
   The explicit evaluation of the determinant of $a_{n,m}$ gives
   \be
      \det (a_{nm}) = 2^{K^2 - K} \prod _{l=0}^{K - 1} \frac{l!}{(K + l)!}
   \ee
   We do recover in this way the factor  $\gamma_K$ (\ref{gammak})(up to a
factor
$2^{K^2 - K}$
due to
   a different normalization normalization of the kernel).

It is quite remarkable that this universal normalizing factor $\gamma_K$
has a geometric interpretation
as a Fredholm determinant of the Dirac Laplacian on the two dimensional
sphere $S^2$. The determinant of the Laplacian has been discussed in the
connection to string theory \cite{Sarnak1,D'Hoker},
and the relation of $\gamma_K$ to this
Fredholm determinant of the Laplacian has been noticed in \cite{ConreyFarmer}.
Indeed let us show that

   \be\label{lap}
    \gamma_K = \frac{e^{K^2( 1 + \gamma )}}{\Delta^{+}(-K)}
   \ee
   where $\gamma$ is Euler's constant and $\Delta^{+} (z)$ the
determinant of a Dirac operator, defined below. The derivation goes as
follows : let
us introduce a function $G(z)$ which satisfies
the functional relation
\be\label{G} G(z+1) = \Gamma (z) G(z).\ee
It is then straightforward to verify that
\be \gamma_K = \prod_{l=0}^{K-1} \frac{l!}{(K+l)!} = 2^{K-2K^2}
\frac{\pi^{K+\frac{1}{2}}}{\Gamma (K+\frac{1}{2})} \left[ \frac
{G(\frac{1}{2})}{G(K+\frac{1}{2})}\right]^2 .\ee
A function $G$, satisfying the functional relation (\ref{G}), is known in
the literature as a Barnes function (or as the inverse of a di-gamma
function).
It is defined by
\be \label{Barnes} G(z+1) = \frac{1}{\Gamma_2(z+1)} = (2\pi)^{z/2} \
e^{-\frac{1}{2}\left[z+(1+\gamma)z^2\right]}
\prod_1^{\infty} \left[ (1+\frac{z}{n})^n e^{-z+z^2/(2n)}\right].\ee
It has been noticed  earlier (\cite{Voros})that this  Barnes function is
related to the Fredholm
determinant of the Laplacian on $S^2$ .
Indeed this Fredholm determinant is the (regularized) product
\be\Delta(z) = \prod_{l} ( 1 - \frac{z}{\la_l})^{g_l}
      \ee
    where the $\la_l$ are the eigenvalues of the Laplacian, and $g_l$ their
degeneracy, i.e. $\la_l = l( l + 1)$
    with multiplicity $g_l = 2 l + 1$, $l = 0,1,2,...$.
    It is convenient to shift $z$ by $1/4$, since this  yields the
    the spectrum of the Dirac operator
    \be
    \sqrt{\lambda_l+ \frac{1}{4}} = l + \frac{1}{2}
    \ee
   Then the regularized (shifted) Fredholm determinant
\be   \Delta(z) = \prod_{l=0}^{\infty} [ (1 - \frac{z}{(l +
     \frac{1}{2})^2})
     e^{\frac{z}{(l + 1/2)^2}} ]^{2 l + 1},
     \ee
  factorizes   as
    \be\label {factorized}
    \Delta(-y^2) = \Delta^{+}(i y) \Delta^{+}(- i y)
    \ee
    with the determinant of the Dirac operator
$\Delta^{+}(z)$  given by \cite{Voros}
    \be
     \Delta^{+}(z) = \prod_{l=0}^{\infty} [ (1 - \frac{z}{l +
     \frac{1}{2}})
     e^{\frac{z}{l + 1/2} + \frac{z^2}{2 ( l + 1/2)^2}}]^{2 l + 1}.
     \ee

    Then this Dirac
determinant $\Delta^{+}$ is related to the Barnes function
by
     \be
     \Delta^{+}(z) = \pi^{-\frac{1}{2}}(2 \pi )^z e^{(1 + \gamma + 2 \log
2)z^2}
     \frac{\Gamma(\frac{1}{2} - z) G(\frac{1}{2} - z)^2}{G(\frac{1}{2})^2}.
     \ee
We thereby recover the  expression relating the moment $\gamma_K$ to the
determinant
     (\ref{lap}).

This relation between the moments of the distribution and the
determinant of the  Dirac operator on $S^2$
 is in fact general. For instance in the simplest case of  a
    single Gaussian  random variable,  the moments are
     \be
       \int_{-\infty}^{\infty} x^{2 K} e^{-x^2} dx = \Gamma( K +
\frac{1}{2}) ;
     \ee
 $\Gamma (K+1/2)$ is thus the equivalent of $\gamma _K$ for this trivial
problem.
     If we consider the "Laplacian", i.e. the harmonic oscillator whose
eigenvalues are $\la_n = n$, then
    the Fredholm determinant $\Delta(\la)$ is
     \ba
     \Delta(\la) &=& -\lambda\prod_{n=1}^{\infty} ( 1 -
\frac{\la}{n})e^{\frac{\la}{n}}\nonumber\\
     &=& \frac{e^{\gamma \la}}{\Gamma(- \la)}
     \ea
     Hence, we have
     \be <x^{2 K}> \frac{e^{\gamma \la}}{\Delta(\la)}\large \vert_{\la = -
(K +\frac{1}{2})} \ee
    The expression (\ref{lap}) is a multi-variable version of this Gaussian
    integral.

An additional point of interest is that the Fredholm determinant of this
Laplacian on $S^2$
 may be factorized further into
    a product of two factors ; it turns out that each factor enters into
the corresponding
expression for the symplectic and  orthogonal cases respectively.
    This will be seen below when we examine the moments related to  the
Bessel kernel.

\section{Bessel kernel}

    We have discussed in our previous work the ensembles invariant under
the unitary symplectic and unitary orthogonal Lie groups \cite{BH1a}.
The kernels
    for those ensembles are \cite{AZ,BHL,Sarnak}
    \be
    K (x,y) = \frac{1}{2\pi} ( \frac{\sin( x - y)}{x - y} \mp
    \frac{ \sin ( x + y)}{x + y})
    \ee
    where the minus sign corresponds to the $Sp$ and the plus sign to
the $O$
    ensemble.
    It is convenient to introduce the Bessel kernel defined by
    \be
    K_{\alpha} (x,y) = \frac{J_\alpha (x) J_{\alpha}^{\prime}(y) -
    J_{\alpha}^{\prime}(x)J_\al (y)}{x - y}
    \ee
    Since $J_{1/2}(x) = \sqrt{2/\pi x} \sin x$, $J_{-1/2}(x) = \sqrt{2/
    \pi x} \cos x$, the kernels for the $Sp$ and $O$ ensembles are both
related to this Bessel kernel
    \be
    K_{\pm}(x,y) = \sqrt{x y} K_{\pm1/2}(x^2,y^2)
    \ee
    namely, $\al = 1/2$ and  $\al = -1/2$ represent respectively  the
    $Sp$ and the $O$ ensemble.
    We consider from now on an arbitary $\al$.
    The 2K-th moment at the origin $(\la = 0)$ is expressed as
    \be
    I_K = \oint \oint \frac{du dv}{(2\pi)^2}
    \frac{\Delta(u^2) \Delta(v^2)}{\prod_{i=1}^K u_i^{2K}
    v_i^{2K}}
    \prod_{i=1}^K (u_i v_i)^\al K_\al(u_i,v_i)
    \ee
    We define now the two functions $\phi(z)$ and $\psi(z)$  by
    \be
    J_\al(\sqrt{z}) = (\frac{\sqrt{z}}{2})^\al \frac{1}{\Gamma(\al +
1)}\phi(z)
    \ee
    and
    \be
    \sqrt{z} J_\al^{\prime}(\sqrt{z}) =
    \frac{z^{\al/2}}{2^\al \Gamma(\al)} \psi
    (z) .
    \ee
    Their expansions in powers of $x$ are given by
    \be
    \phi(x) = \sum_{n=0}^{\infty}
    \frac{(-1)^n x^n}{4^n n! \prod_{l=1}^n (\al + l)}
    \ee
    \be
    \psi(x) =  \sum_{n=0}^{\infty} \frac{(-1)^n x^n (\al + 2n)}{4^n n!
    \prod_{l=0}^n (\al + l)}
    \ee
    Keeping aside trivial factors we are then led to the  kernel
    $\tilde K_\al(x,y)$  defined as
    \be
      \tilde K_\al(x,y) = \frac{1}{2(x - y)} [\phi(x)\psi(y) -
\psi(x)\phi(y)]
      \ee
  As before, we have
   \be
   I_K = \det(a_{nm})
  \ee
  with
 \be
     a_{nm} = \frac{1}{n! m!} \frac{\partial^n}{\partial u^n}
     \frac{\partial^m}{\partial v^m} \tilde K_\al(u,v)|_{u=v=0}
 \ee
 This determinant may be computed  explicitly, and it is given by
 \be
  I_K = 4^{-K^2 -\al K} \prod_{l=0}^{2K - 1} \frac{1}{(\al + l)!}
  \ee
  (We have $I_1 = \frac{1}{4}, \frac{1}{3 \pi}, \frac{1}{\pi}$ for
   $\al = 0,\frac{1}{2}$ and $-\frac{1}{2}$, respectively.)

   It is interesting to relate the three determinants that we have
introduced hereabove for
the sine-kernel and for the
$Sp$ and $O$ cases. The determinant for the  sine-kernel (\ref{sinedet}) is
   \ba\label{iu}
   I_U = \det \left(\matrix{1&0&-\frac{1}{6}&0&\ldots\cr
   0&\frac{1}{3}&0&-\frac{1}{30}&\ldots\cr
   -\frac{1}{6}&0&\frac{1}{20}&0&\ldots\cr
   0&-\frac{1}{30}&0&\frac{20}{7!}&\ldots\cr}\right).
   \ea
    In the symplectic case, $\al = \frac{1}{2}$, we have
    \ba\label{isp}
    I_{Sp} = \det\left(\matrix{\frac{1}{3}& - \frac{1}{30}&\ldots\cr
    -\frac{1}{30}& \frac{20}{7!}&\ldots\cr
    \ldots&\ldots&\ldots\cr}\right).
    \ea
    In the orthogonal case, the determinant becomes for $\al= -\frac{1}{2}$,
    \ba\label{io}
    I_{O} = \det\left(\matrix{1& - \frac{1}{6}&\ldots\cr
    -\frac{1}{6}& \frac{1}{20}&\ldots\cr
    \ldots&\ldots&\ldots\cr}\right).
    \ea
    Thus, we find the factorization of (\ref{iu}) as the product of
(\ref{isp}) and
    (\ref{io}), up to a trivial numerical factor due to the normalizations,
    \be
      I_U = I_{Sp} \times I_{O}
    \ee
    The factors $\gamma_K$ for the unitary, symplectic and orthogonal
    case are related as $2^{K^2 - 1}\gamma_K^{(U)} = \gamma_K^{(Sp)}
    \gamma_K^{(O)}$, and $\gamma_K^{(U)} = (\prod_{l=1}^{K-1}l!)^2/(
    \prod_{l=1}^{2K-1}l!), \gamma_K^{(Sp)} = 2^{K(K+1)/2}\prod_{l=1}^K
    l!/\prod_{l=1}^K (2l)!$ and $\gamma_K^{(O)} 2^{K(K+1)/2 - 1}
     \prod_{l=1}^{K-1} l!/\prod_{l=1}^{K-1}(2l)!$.
     It is again remarkable that, for  arbitrary $\al$, $\gamma_K$ may
still be expressed as the Fredholm determinant
 of the Laplacian, in which the eigenvalues are shifted by the amount $\al$
     \cite{Voros}.

\section{Airy kernel}
    When the eigenvalues lie near an edge $\la_c$ of the support of the
asymptotic density of states (an edge
of Wigner's semi-circle in the Gaussian case), in a neighbourhood of size
$\displaystyle N^{-2/3}$ of that edge, there is a cross-over
 from the sine-kernel to the Airy kernel.
    In terms of the Airy function $A_i(x)$,  defined by
\be\label{Airy}
     A_i(x) = \frac{1}{2\pi} \int_{-\infty}^{\infty} dz e^{\frac{i}{3} z^3 +
i z x},
\ee
which satisfies the differential equation
\be
    A_i^{\prime\prime}(x) = x A_i(x),
\ee
one has
\be \label{Airykernel}
    K(x,y) = \frac{A_{i}(x) A_i^{\prime}(y) - A_i^{\prime}(x) A_i(y)}{x - y}.
\ee
In (\ref{Airykernel}) we have used the scaling variables $x$ and $y$
proportional
to $\displaystyle N^{2/3}(\la-\la_c)$ .

  There are two ways to obtain the moments under
consideration. The first one is to write as before

\be
     I_K = < [\det(\la_c - X)]^{2K} > = \oint \frac{du}{2\pi i}
\frac{\Delta(u)\Delta(v)}{\prod_{i=1}^K
u_i^K v_i^K
} \prod_{i=3D1}^K K(u_i,v_i)
\ee
but  in this case, there are three periodic structure due to three
valleys of Airy functions, and
the result is more complicated. It does not seem to be expressible as simple
products of  gamma-functions.
However we can use a direct method starting with the expression
\be\label{Airy1}
     I_K = < [\det(\la_c - X)]^{2K} > = \frac{1}{(2\pi)^{2K}}
     \int_{-\infty}^{\infty} dz
\Delta^2(z) e^{\frac{i}{3}\sum_{i=1}^{2K} z_i^3}
\ee
    This representation is the  edge limit   $\la_l\rightarrow 0$  of
\be
   F_{2K} = \int_{-\infty}^{\infty}\prod dz_l \oint \frac{du_i}{2\pi i}
\frac{\Delta(z)\Delta(u)}
{\prod_i\prod_l(u_i-\la_c + \la_l)}\ e^{N\displaystyle
\sum_1^{2K}(\frac{i}{3} z_l^3 + i  z_l u_l)}
\ee
The sums and products over $l$ run from $l=1$ to $l=2K$.
The dependence of $F_{2K}$ on $N$ is of order $N^{\frac{2}{3}K^2 - K}$.

We may then use the standard orthogonal polynomial method.
To the complex measure
\be
    d\mu = dz e^{\frac{i}{3}z^3}
\ee
we associate the orthogonal polynomials $p_n$  defined as
\be
   p_n(x) = x^n + \hbox{lower degree},
\ee
and
\be
   \int d\mu p_n(x) p_m(x) = h_n \delta_{n,m}
\ee
The integral of (\ref{Airy}) is then simply
\be
   I = K! h_0 h_1 \cdots h_{K-1}
\ee
Note that this looks  similar to the partition function of a matrix model,
but here it
is the partition function of a $K\times K$ matrix model, instead of
$N\times N$ ($K$ is finite, since it is the order of the moment that we are
considering, whereas
$N$ goes to infinity). Therefore this is for any K a completely explicit
expression of the moments at the edge.
Those coefficients $h_n$ are expressible in terms of ratios of determinants
constructed with the
moments of the measure :
\be h_n = \frac {d_n}{d_{n-1}} \ee
with
\ba\label{d}
   d_n = \det \left(\matrix{m_0&m_1&\cdots&m_n\cr
   m_1&m_2&\cdots&m_{n+1}\cr
   \cdots&\cdots&\cdots&\cdots\cr
   m_n&m_{n+1}&\cdots&m_{2n}}\right)
   \ea
in which the $m_n$ are the moments of the measure. Those determinants are
constants along anti-diagonal lines (Hankel
determinants). Then the
product
$h_0h_1\cdots h_{K-1}$ is reduced to  a single determinant.
For example, we have for $K=4$
\ba\label{Airy2}
   h_0h_1h_2h_3 = \det \left(\matrix{C_1&-i C_2&0&iC_1\cr
   -iC_2&0&iC_1&2C_2\cr
   0&iC_1&2C_2&0\cr
   iC_1&2C_2&0&-4C_1}\right).
 \ea
with $C_1 =  A_i(0) = 3^{-2/3}/\Gamma(2/3)$, $C_2= A_i^{\prime}(0) =
- 3^{-1/3}/\Gamma(1/3)$,
since all the moments up to $m_6$ are easily expressible in terms of $m_0$
and $m_1$ alone.

More generally we have
\be\label{moments}
    m_n = \int z^n d\mu = (-i)^n ( n- 2)(n - 5) (n - 8)\cdots \tilde A_n
\ee
where $\tilde A_n = C_1$ for n=0 (modulo 3), and $\tilde A_n = C_2$
 for n=1 (modulo 3) and $\tilde A_n = 0$ for n=2 (modulo 3).
The last parenthesis of the product in the r.h.s. of (\ref{moments}) is the
rest of the division of $n-2$ by 3.
Then, $d_n$ is the determinant of a Hankel matrix, whose matrix
elements
in the first row are $ [<z^0>,<z>,<z^2>,\cdots] =
[C_1,-iC_2,0,iC_1,2C_2,0,-4C_1,10 i C_2,0,
-28i C_1,-80 C_2,0,\cdots]$, and all the others are given by the Hankel
rule.
In this way we obtain sucessively,
\ba\label{Hankel2}
   h_0 &=& C_1 = 0.355028053\nonumber\\
   h_0h_1 &=& C_2^2 = 0.066987483\nonumber\\
   \prod_{0}^2 h_l &=& 2 C_2^3 + C_1^3 = 0.010074161\nonumber\\
   \prod_{0}^3 h_l &=& - 8 C_1 C_2^3 - 3 C_1^4 = 0.001580882\nonumber\\
   \prod_{0}^4 h_l &=& 72 C_2^5 + 28 C_1^3 C_2^2 = 0.000313095517\nonumber\\
   \prod_{0}^5 h_l &=& -2160C_2^6 - 1952 C_1^3 C_2^3 - 432 C_1^6
= 0.000090756324
   \ea
Therefore for the edge problem we have found moments given by $\gamma_K$ 's
which are more complicated
since $\gamma_K = \prod_{0}^{2K-1}h_l$.
The result is explicit for any finite K, but we have
not succeeded to
continue it to non-integer K. The numerical values indicate a smooth curve
in a logarithmic plot.

\section{Finite N results}

   We have derived in our previous paper \cite{BH1a}
the correlation functions  of the characteristic polynomials
    under the form of a determinant.
  \ba \label{F}&& F_K({\la}_1,\cdots,{\la}_K) = \langle \prod_1^{K}
\det(\la_l -X) \rangle
\nonumber\\ = &&
\frac{1}{\Delta(\la_1,\cdots,\la_K)}\det \left| \begin{array}{clcr}
p_M(\la_1) &p_{M+1}(\la_1)&\cdots&p_{M+K-1}(\la_1)
\\p_M(\la_2) &p_{M+1}(\la_2)&\cdots&p_{M+K-1}(\la_2)\\ \vdots\\p_M(\la_K)
&p_{M+1}(\la_K)&\cdots&p_{M+K-1}(\la_K)\end{array} \right|,\ea
in which $X$ is an $M\times M$ random matrix.

The polynomial $p_n(x)$ are the (monic) orthogonal polynomials,
 whose coefficients of highest
degree are equal to unity
\be p_n(x) = x^n + \rm{lower degree}.\ee

If we are concerned simply with the moments of the distribution of a single
characteristic polynomial, we obtain from (\ref{F})
\ba\label{F1}
 \mu_K(\la) =&& F_K({\la},\cdots,{\la}) = \langle \ [\det(\la-X)]^K\
\rangle\nonumber\\ =&& \frac{(-1)^{K(K-1)/2}}{\prod_{l=0}^{K-1} (l!)}
\det
\left|
\begin{array}{clcr} p_M(\la) &p_{M+1}(\la)&\cdots&p_{M+K-1}(\la)
\\p'_M(\la) &p'_{M+1}(\la)&\cdots&p'_{M+K-1}(\la)\\
\vdots\\p^{(K-1)}_M(\la)
&p^{(K-1)}_{M+1}(\la)&\cdots&p^{(K-1)}_{M+K-1}(\la) \end{array}\right|.\ea

For the Gaussian distribution,
\be \label{GAUSS} P(X) = \frac{1}{Z_M} \exp - \frac{N}{2} \rm Tr X^2 ,\ee
with
\be M= N-K,\ee
the polynomial $p_n(x)$ are the
Hermite polynomials $H_n(x)$, defined with our normalization as
\be\label{Hermite}
 H_n(x) = \frac{(-1)^n}{N^n} e^{Nx^2/2} (\frac{d}{dx})^n e^{-Nx^2/2} =
x^n + \rm{l.d.}.\ee

The integral representation
\be\label{contour} H_n(x) = \frac{(-1)^n n!}{N^n}\oint \frac{dz}{2i\pi}
\frac{e^{-N(z^2/2+xz)}}{z^{(n+1)}}\ee
over a contour which circles around the origin in the z-plane, turns out to
be well adapted.

Note that all these expressions are all valid for finite N.
We may thereby recover readily  several results that we have discussed in
the previous sections. For instance, let us assume that $M$ is an even number,
and consider the center
value $\la = 0$ (since the
dependence in $\la$ is known to be
 contained entirely in the overall factor $[\r(\la)]^{K^2}$, as far as the
coefficient
$\gamma_K$ is concerned,
it is sufficient to put simply $\la=0$).

The Hermite polynomials $H_n(x)$ vanish for odd n at $x=0$. Similarly the
odd derivatives
of $H_n(x)$ for even n,  also vanish at $x=0$. Hence,
the elements of the determinant (\ref{F1}) are alternatively non-zero then
zero.
Thus the determinant is decomposed into a product of two determinants ;
this is  the exact
 phenomenon for N finite of
the factorization of the symplectic and  orthogonal
determinants that we have seen earlier for large N.
Since the matrix elements of (\ref{F1}) at $\la=0$ are all expressed as
derivatives of Hermite polynomials at the origin,
it is possible to compute this determinant exactly for finite and  arbitrary
 M and K .
For the  even M case,
\ba
  F_{2K}(0) &=& \frac{(-1)^{K(2K-1)}}{\prod_{l=3D0}^{2K-1} (l!)}
  \det\left|
\begin{array}{clcr} H_M(0) &H_{M+2}(0)&\cdots
\\H''_M(0) &H''_{M+2}(0)&\cdots\\
\vdots\\H^{(2K-2)}_M(0)
&H^{(2K-2)}_{M+2}(0)&\cdots \end{array}\right|
\nonumber\\
&\times& \det\left|
\begin{array}{clcr} H'_{M+1}(0) &H'_{M+3}(0)&\cdots
\\H'''_{M+1}(0) &H'''_{M+3}(0)&\cdots\\
\vdots\\H^{(2K-1)}_{M+1}(0)
&H^{(2K-1)}_{M+3}(0)&\cdots \end{array}\right|
\ea
We denote each determinant as $ I^{(1)}/N^{KM/2}$ and $ I^{(2)}/N^{KM/2}$
respectively, and
\be\label{F2K}
   F_{2K}(0)=
  I^{(1)} I^{(2)} \frac{1}{N^{KM}}
  \frac{1}{\prod_{l=0}^{2K-1}l!}.
\ee
The  above two determinants are easily computed through the explicit
expressions for  the
$H_n(x)$'s,
\be
   H_{2n}(x) = \frac{1}{n} (-1)^n (2n-1)!! \sum_{m=0}^{\infty}
\frac{(-n)(-n+1)\cdots
(-n+m-1)}{(\frac{1}{2})(\frac{1}{2}+ 1)\cdots (\frac{1}{2} + m - 1)}
\frac{1}{m!}
(\frac{N x^2}{2})^m,
\ee
\be
  H_{2n+1}(x) = \frac{1}{N^n}(-1)^n (2n+1)!! x \sum_{m=0}^{\infty}
\frac{(-n)(-n+1)\cdots
(-n+m-1)}{(\frac{3}{2})(\frac{3}{2}+ 1)\cdots (\frac{3}{2} + m - 1)}
\frac{1}{m!}
(\frac{N x^2}{2})^m.
\ee
The two determinants contain overall products of factors of the form $(2n
-1)!!$ ; once
they are extracted
one finds
\ba\label{I1}
    I^{(1)} &=& C(M + 2K - 3)!! ( M + 2 K - 5)!! \cdots ( M -
1)!!\nonumber\\
           &=& C \frac{1}{2^{\frac{(K(M+K-3)}{2}}}\prod_{l=1}^K
   [\frac{\Gamma(M + 2l -2)}{\Gamma(\frac{M}{2} + l -1) }]
 \ea
 \ba\label{I2}
  I^{(2)} &=& C(M + 2K -1)!! (M + 2K - 3)!! \cdots (M + 1)!!\nonumber\\
   &=& C \frac{1}{2^{\frac{K(M+K-1)}{2}}}\prod_{l=1}^K
   [\frac{\Gamma(M + 2l)}{\Gamma(\frac{M}{2} + l)}]
 \ea
with
\be
C = 2^{\frac{K(K-1)}{2}}\prod_{l=0}^{K-1} l!
\ee
which is independent of $M$.
   In the large M limit,  from the Stirling formula, we have
    \be
       I^{(1)} \simeq C M^{\frac{MK + K(K-1)}{2}} e^{- \frac{MK}{2}}
2^{\frac{K}{2}}
   \ee
   \be
       I^{(2)} \simeq C M^{\frac{MK + K(K+1)}{2}} e^{- \frac{MK}{2}}
2^{\frac{K}{2}}
   \ee
It is remarkable that, even for finite M (M is the size of the random
matrix),
$F_{2K}(0)$ for this Gaussian distribution, already exhibits the  factor
$\gamma_K = \prod_{l=0}^{K-1} l!/((K +
l)!
= [\prod_{l=0}^{K-1}l!]^2/\prod_{l=0}^{2K - 1}l!$, which is known to be
universal in the large M-limit.
It is indeed  obtained from the product of the factor $C$ and
$1/(\prod_{l=0}^{2K-1}
l!)$ in (\ref{F2K}).
This means that at each order of the $1/N $ expansion, we keep this
universal
factor  for $F_{2K}(0)$.
     In the large N limit (M = N $-$ K), $F_{2K}(0)$ becomes
 \be\label{F0}
    F_{2K}(0) \simeq (2 N)^{K^2} e^{-N K} \prod_0^{K-1} \frac{l!}{(K + l)!}.
 \ee
 In the previous paper, we have derived $F_{2K}(\la)$, in the large N limit,
as
 \be\label{F0X}
   F_{2K}(\la,\cdots,\la) \simeq (2 \pi \rho(\la)N)^{K^2}
   e^{-N K} \prod_0^{K-1} \frac{l!}{(K + l)!}
 \ee
 At the band center, $\la = 0$, the density of state is
 $\rho(0) = \frac{1}{\pi}$
 for the Gaussian distribution. Therefore, (\ref{F0}) is indeed consistent
with
(\ref{F0X}).

 It may be interesting to note that one of the factors  of (\ref{F0X}),
namely $\prod_{l=0}^{2K - 1}(l!)$,
appears in  $F_{2K}(\la)$ in (\ref{F1}) . This factor, a product of
  gamma-functions, remains for any set of  orthogonal
  polynomials, since it stands in the front of the determinant of
(\ref{F1}).

The factor $e^{-NK}$ is cancelled by the normalization \cite{BH1a}. For
$\la\neq 0$, we have evaluated $F_{2K}(\la)$. We have here considered the
finite N
case
to see the universal factor $\gamma_K$.

One can recover again the Airy limit by the use of eq.(\ref{F1}).
   We use once more the properties of the Hermite polynomials such as
    \be\label{successive}
      H_n'(x) = n H(x)
    \ee
    and their explicit integral representation
    \be
      H_n(x) = \frac{1}{\sqrt{2 \pi}} N^{\frac{1}{2}} e^{\frac{N}{2}x^2}
      \int_{-\infty}^{\infty} ds s^n e^{-\frac{N}{2} s^2 - i N x s} .
    \ee
    We set $n = \delta + N$, and after exponentiation, we have
    \be
      H_n(x) =  \frac{1}{\sqrt{2 \pi}} N^{\frac{1}{2}} e^{\frac{N}{2}x^2}
      \int_{-\infty}^{\infty} ds s^{\delta} e^{- N f(s)}
  \ee
  where $f(s) = \frac{1}{2} s^2 + i  s x - \log s$.
  The saddle points are degenerate at the edge $x = 2$. The vicinity of
this point
is blown out through a change of variables, with a  scaling ansatz,
  \be
     x = 2 + N^{-\alpha} y
   \ee
and
   \be
      s = - i + N^{-\beta} z
   \ee
     If one expands $f(s)$ up to order $z^3$, one sees that
  in the proper scaling choice $\alpha = 2/3$ and
   $\beta = 1/3$, one recovers the Airy limit which governs the properties
 of
the system in a
neighbourhood of size $N^{-2/3}$ of the edge of Wigner's semi-circle. Then,
the integral becomes
   \be
     I = (-i)^{\delta} N^{-\frac{1}{3}} \int_{-\infty}^{\infty}
     dy e^{\frac{i}{3}y^3 + i z y}
   \ee
      This is indeed the Airy function $A_i(z)$ of (\ref{Airy}).
   \be
      H_{N + \delta} (x) = \sqrt{2 \pi N} e^{2N} (-i)^{\delta}
      A_i((x - 2) N^{\frac{2}{3}})
   \ee
   We now consider all the $\la_i = 2$, and the determinant  (\ref{F1})
becomes in the large N limit
   a determinant of Airy functions.
   If we replace  $H_{M + 2K -1}$ at
   the right-up corner of the determinant by  the Airy function $A_i(0)$,
   the other matrix elements become  derivatives of the Airy function,
   since there is a the recursion relation  (\ref{successive}).
   For example,in the  $K=1$ case, we have
   \be
   \det\left|
\begin{array}{clcr} H_{M}(2) &H_{M+1}(2)
\\H_{M}'(2) &H_{M+1}'(2)
 \end{array}\right| \sim
\det\left|
\begin{array}{clcr} \frac{N^{\frac{2}{3}}}{M + 1} A_i'(0) & A_i(0)
\\ \frac{N^{\frac{4}{3}}}{M + 1} A_i''(0)& N^{\frac{2}{3}}A_i'(0)
 \end{array}\right|\ee

  Then, we find in the large N limit, with $N = M - K$,
  \be
  F_{2K}(2) = \frac{N^{\frac{2}{3}K(K+1)}}{\prod_{l=0}^{2K-1} l!}
 \det\left|
\begin{array}{clcr} \cdots & A_i'(0)& A_i(0)
\\\cdots& A_i''(0)& A_i'(0)\\
\cdots& \cdots &\cdots \end{array}\right|
\ee
The above determinant was discussed earlier. Note the factor
$1/\prod_{l=0}^{2K - 1} l!$  in front.

\section{Derivative moments}

   The same techniques may also  be used if one is interested in the
moments of the D-th derivatives
   (D = 1,2,...) of the characteristic polynomials. Let us consider for
instance
   \be
     F_{2K}^{(D)}(\la_1, \cdots,\la_{2K}) = < \prod_{l=3D1}^{2K}
     \frac{\partial^D}{\partial \la_i^D}\det ( \la_i - X) >
   \ee
   From (\ref{F1}), one sees immediately that it has also the form of a
determinant :
   \ba\label{DF1}
    F_{2K}^{(D)}({\la}_1,\cdots,{\la}_{2K}) =
\frac{1}{\Delta(\la_1,\cdots,\la_{2K})}\det \left| \begin{array}{clcr}
p_M^{(D)}(\la_1) &p_{M+1}^{(D)}(\la_1)&\cdots&p_{M+2K-1}^{(D)}(\la_1)
\\p_M(\la_2)^{(D)} &p_{M+1}^{(D)}(\la_2)&\cdots&p_{M+2K-1}^{(D)}(\la_2)\\
\vdots\\p_M^{(D)}(\la_{2K})
&p_{M+1}^{(D)}(\la_{2K})&\cdots&p_{M+2K-1}^{(D)}(\la_{2K})\end{array}
\right|.\nonumber\\
\ea

 When all the $\la_i$'s are equal, we have
   \ba\label{F3}
 F_{2K}^{(D)}({\la},\cdots,{\la}) &=& \langle \ [\frac{d^D}{d \la^D}
 \det(\la-X)]^{2K}\
\rangle\nonumber\\ &=& \frac{(-1)^{K(2K-1)}}{\prod_{l=0}^{2K-1} (l!)}
\det
\left|
\begin{array}{clcr} p_M^{(D)}(\la)
&p_{M+1}^{(D)}(\la)&\cdots&p_{M+2K-1}^{(D)}
(\la)
\\p_M^{(D+1)}(\la) &p_{M+1}^{(D+1)}(\la)&\cdots&p_{M+2K-1}^{(D+1)}(\la)\\
\vdots\\p^{(D+2K-1)}_M(\la)
&p^{(D+2K-1)}_{M+1}(\la)&\cdots&p^{(D+2K-1)}_{M+2K-1}(\la)
\end{array}\right|.
\nonumber\\
\ea

If we set
$\la=0$ it may be  again decomposed into  a product of two determinants .
Let us  assume
for definiteness that both $M$ and $D$ are even.
Then, we have
 \be
 I^{(1)} = \det\left|
\begin{array}{clcr} H_M^{(D)}(0) &H_{M+2}^{(D)}(0)&\cdots
\\H_M^{(D+2)}(0) &H_{M+2}^{(D+2)}(0)&\cdots\\
\vdots\\H^{(D + 2K-2)}_M(0)
&H^{(D+2K-2)}_{M+2}(0)&\cdots \end{array}\right|
\ee
\be
I^{(2)} = \det\left|
\begin{array}{clcr} H_{M+1}^{(D+1)}(0) &H_{M+3}^{(D+1)}(0)&\cdots
\\H_{M+1}^{(D+3)}(0) &H_{M+3}^{(D+3)}(0)&\cdots\\
\vdots\\H^{(D + 2K-1)}_{M+1}(0)
&H^{(D+2K-1)}_{M+3}(0)&\cdots \end{array}\right|
\ee
Using the explicit expressions for the Hermite polynomials , we
can compute these
determinants.
We find for arbitrary M,D and K,
\be
  F_{2K}^{(D)}(0) = \frac{1}{N^{K(M-D)}}
I^{(1)}I^{(2)}\frac{1}{\prod_{l=0}^{2K-1}
l!}
\ee
\ba\label{DI1}
I^{(1)} &=& (M + 2K - 3)!! (M + 2K - 5)!! \cdots ( M - 1)!!\nonumber\\
        &\times& \prod_{l=0}^{K-1}[(\frac{M}{2}+l)(\frac{M}{2}+l-1)\cdots
        (\frac{M}{2} - \frac{D}{2} + l
        + 1)]\nonumber\\
        &\times& 2^{\frac{DK + K(K-1)}{2}}\prod_{l=0}^{K-1} l!
\ea
\ba\label{DI2}
I^{(2)} &=& (M + 2K - 1)!! (M + 2K - 3)!! \cdots ( M + 1)!!\nonumber\\
        &\times& \prod_{l=0}^{K-1}[(\frac{M}{2}+l + 1)(\frac{M}{2}+l)\cdots
        (\frac{M}{2} - \frac{D}{2} + l
        + 2)]\nonumber\\
        &\times& 2^{\frac{DK + K(K-1)}{2}}\prod_{l=0}^{K-1} l!
\ea

(One may easily check these results for D = M, since
the matrix elements below the diagonal vanish, i.e.
the determinants are  then simply given by the product of the diagonal
elements,
$\prod_{l=0}^K(M + 2l)!$ which agrees with  (\ref{DI1}).
When $D=0$, it reduces to the previous expression (\ref{I1}).
$I^{(2)}$ is obtained from $I^{(1)}$ by the shift $M\rightarrow M +2$.)

In the large N limit, we have
\be
  F_{2K}^{(D)}(0) \simeq (2 N)^{K^2 + 2KD} e^{- KN} \frac{1}{2^{2KD}}
  \prod_{l=0}^{K-1}\frac{l!}{(K + l)!}
  \ee
Hence, for this derivative moments at finite M,
again
 the universal factor $\gamma_K$ is present, and it persists of course in
 the large N limit.

These results lead to the conjecture that the average of the moment of
derivatives of
the Riemann
zeta-function along the critical line
\be
   I = \frac{1}{T}\int_{0}^T dt |\frac{d^D}{d t^D} \zeta(\frac{1}{2} +
it)|^{2K},
\ee
   also have this universal
factor $\gamma_K$.

\section{External source}

   We now consider the case in which
    the external source matrix $A$ is coupled to the random matrix $X$.
The measure of the random matrix $X$ is
\be
  d\mu(X) =\frac{1}{Z}e^{-\frac{N}{2}\tr X^2 + N \tr X A} d^{N^2}X
\ee
The eigenvalues of the matrix $A$ are denoted by $a_i$, $i=1,\cdots,N$.
In such cases, the standard orthogonal polynomial method cannot be used.
However, the n-point correlation functions $\rho(\la_1,\cdots,\la_n)$
have been
found to be described again by the determinant of a kernel ;  from there
 the level spacing
probability $p(s)$ has been also investigated \cite{BH3}.
If we specialize to a source which has two opposite eigenvalues, namely
$a_i = +a$ for
$i=1,\cdots,N/2$ and $a_i = -a$ for $i= N/2+1,\cdots,N$,  one finds a support
for the eigenvalues made of two disconnected segments for $a>1$. If one
tunes
the external source so that $a=1$, i.e. $a_i=\pm 1$, the gap between the
two segments closes
and the spectrum consists of a single segment for $a<1$.  We want to
investigate here the critical point $a=1$
which gives rise to yet another class of universality.

The
moments $F_{2K}(\la,\cdots,\la)$ at $\la = 0$ at this closing
gap point may turn out to have interesting applications.

Since $X$ and $A$ are Hermitian matrices,  we
write
\be
\tr X A = \tr U^{-1} X_{0} U A_{0}
\ee
where $X_{0} = diag(x_1,\cdots,x_M), A_{0} = diag(a_1,\cdots,a_M)$, and
$U$ belongs to the unitary group.
The integration  over this unitary group $U$ is well known from the work of
Harish-Chandra,
Itzykson-Zuber
\cite{Harish-Chandra,Itzykson-Zuber}, and this is the starting point of the
formulae
found in \cite{BH2}. For instance the n-point
correlation
functions are given by the determinant of the $n\times n$ matrices  made
with the
kernel $K(\la_i,\la_j)$ with
\be
K(\la,\mu) = \int_{-\infty}^{\infty} \frac{dt}{2\pi}\oint
\frac{du}{2\pi i}
\prod_{l=1}^{N} \frac{a_l - i t}{u - a_l} \frac{1}{u - it}
e^{-\frac{N}{2}t^2 + N i t \la - \frac{N}{2} u^2 + N u \mu +
\frac{N}{4}\la^2 -
\frac{N}{4}\mu^2}
\ee
where the contour encloses all the $a_l$'s.

However we may proceed without that here and compute the correlation
functions of
the characteristic polynomials directly.
Indeed
\ba
F_K(\la_1,\cdots,\la_K) &=& < \prod_{\al = 1}^K \det(\la_\al - X) >
\nonumber\\
&=& \frac{1}{Z}\int dX \prod_{\al=1}^{K}\det(\la_{\al} - X)
e^{-\frac{N}{2}\tr
X^2 + N \tr X A}
\ea
In the above equation, the random matrix $X$ is assumed to be
an $M\times M$ matrix, with $M = N - K$, as before. When $K=1$, this
gives a polynomial, which was investigated before \cite{Zinn-Justin}.

The explicit integration over the unitary group
\cite{Harish-Chandra,Itzykson-Zuber}, leads
to
\be \label{source}
F_{K}(\la_1,\cdots,\la_K) = \int \prod_{i=1}^M dx_i
\frac{\Delta(x_1,\cdots,x_M;\la_1,\cdots,\la_K)}
{\Delta(a)\Delta(\la)} e^{-\frac{N}{2} \sum_{i=1}^M x_i^2 + N \sum_{i=1}^M
 x_i a_i}
\ee
where $\Delta(x_1,\cdots,x_M;\la_1,\cdots,\la_K)$ is the Van der Monde
determinant $(M+K)\times (M+K)$ made with the $x$'s and the $\la$'s.
This determinant may be replaced by a determinant of (monic) polynomials ,
and  we choose
the Hermite polynomials
defined in (\ref{Hermite}). It is then straightforward to verify that

\be
\int_{-\infty}^{\infty} H_n(x) e^{-\frac{N}{2}x^2 + N a x}dx =
a^{n}e^{\frac{N}{2}a^2}
\sqrt{\frac{2\pi}{N}}
\ee
Therefore we can explicitely integrate over the M variables $x_i$'s
in (\ref{source}) and  one obtains
\ba \label{mixed}
&&F_K(\la_1,\cdots,\la_K) =
\frac{1}{\Delta(a)\Delta(\la)}\nonumber\\
&&\times\det \left|
\matrix{ 1
&\cdots&1&H_0(\la_1)& \ldots&H_0(\la_K)\cr
\vdots&\ddots&\vdots&\vdots&\ddots&\vdots\cr
a_1^{M+K-1}&\ldots&a_M^{M+K-1}&H_{M+K-1}(\la_1)&\ldots&H_{M+K-1}(\la_K)\cr
}\right|.
\ea

Let us first check that  in the limit of a vanishing source
in which all the $a_i\rightarrow 0$, we do recover the previous formula
(\ref{F}).
The column which depends upon $a_i$ is expanded in Taylor series
around $a_1$, and subtracting the successive columns, we
obtain, after factoring the  Van der Monde determinant $\Delta(a)$ which
cancels the denominator,
a vanishing upper triangle (up to the M-th column) , ones on
the diagonal  and powers of the $a_i$'s below the diagonal. We can now let
the $a_i$'s go to zero
and we are left with the $K \times K$ of (\ref{F}). (In \cite{BH1a} we gave
a different derivation of this same formula).

If we return to an arbitrary non vanishing external source, we
may proceed by returning to (\ref{mixed}) and define $G_K(b_1,\cdots,b_K)$,
\ba
G_K(b_1,\cdots,b_K)&=& \int F_K(\la_1,\cdots,\la_K)\Delta(\la)
e^{-\frac{N}{2}\sum\la_l^2+ N \sum
\la_l b_l} \prod d\la_i\nonumber\\
&=& \frac{\Delta(a;b)}{\Delta(a)}
e^{ \frac{N}{2}\sum b_l^2}
\ea

We may now recover $F_K$ by taking the Fourier transform of
$G_K(ib_1,\cdots,ib_K)$,
\be
\int G_K(ib_1,\cdots,ib_K) e^{-iN\sum \la_i b_i}\prod_{i=1}^K
\frac{db_i}{2\pi}
= (\frac{1}{N})^K \Delta(\la)F_K(\la_1,\cdots,\la_K) e^{-\frac{N}{2}\sum
\la_l^2}
\ee
Therefore, we obtain the following explicit formula,
\ba\label{externalmoment}
F_K(\la_1,\cdots,\la_K) &=&
 \frac{N^K}{\Delta(\la)} e^{\frac{N}{2}\sum \la_l^2}\frac{1}{K!}\nonumber\\
&\times&
\int \prod_{i=1}^K \frac{db_i}{2\pi}\prod_{j=1}^M (ib_l - a_j)\prod_{l<l'}^K
 (ib_l - ib_{l'})
e^{-\frac{N}{2}\sum b_l^2}\det( e^{-iN\la_l b_{l'}})\nonumber\\
 \ea
Note that we could replace in the integrand of (\ref{externalmoment})
$\det( e^{-iN\la_l b_{l'}})$  by the diagonal term
$\displaystyle e^{-iN\sum_1^K\la_l b_l}$ and cancel the
$K!$ in the denominator.
Again we can examine the limit of this formula when the external source
goes to zero, and putting all $\la_i=\la$, we
obtain
\be\label{previous}
F_{2K}(\la) = \frac{N^{K(2K+1)}}{\prod_{l=0}^{2K-1} l!}\frac{1}{(2K)!}
e^{KN\la^2}\int
\prod_{l=1}^{2K} b_l^M \Delta^2(b)e^{-\frac{N}{2}\sum b_l^2 - iN\la \sum b_l}
\prod_{l=1}^{2K}\frac{db_l}{2\pi},
\ee
(we have considered $F_{2K}$ instead of $F_K$ in order to compare with our
previous results).
In the large N limit, we exponentiate $b_l^M$, ($M = N-K$), and look for
the saddle
points which are the roots of the equation $b^2 + i \la b - 1=0$ ; let us
call the two roots
$b^{+}$ and $b^{-}$.
The difference  $|b^{+}-b^{-}| = 2 \pi \rho(\la)$.
The leading saddle-point for the  $b_l$'s,$(l=1,\cdots,2K), $ is obtained by
choosing  half of them
equal to $b^{+}$,
 and $b^{-}$ for the another half. The following Gaussian
integral with a Van der Monde determinant,
\be\label{formula}
\frac{1}{K!}\int \prod_{i=1}^K db_i e^{-\frac{N}{2}f^{''}b^2}
\prod_{i<j}^K (b_i - b_j)^2 =
(\frac{2\pi}{Nf^{''}})^{\frac{K}{2}}
\frac{\prod_{l=0}^{K-1}l!}{(Nf^{''})^{\frac{K(K-1)}{2}}},
\ee
where $f^{''}$ is the second derivative of $f$ at the saddle-point, allows
us to complete the calculation.
Integrating then  around the
saddle-points $b^{+}$ and $b^{-}$, and
keeping in mind the combinatorial factor $\frac{(2K)!}{K!K!}$, which is
the number of choices of K $b^{+}$ and  K
$b^{-}$ among the $2K$  $b_l$'s,
 we recover precisely our previous result,
\be
e^{-NKV(\la)}F_{2K}(\la) = (2\pi N \rho(\la))^{K^2} e^{-NK}
\gamma_K
\ee
where $\gamma_K = (\prod_{l=0}^{K-1}l!)^2/(\prod_{l=0}^{2K -1}l!) =
(\prod_{l=0}^{K-1}l!)/\prod_{l=0}^{K -1}(K+l)!$, and $V(\la)=
\frac{\la^2}{2}$.

From the expression  (\ref{externalmoment}), it is also easy to obtain
the moments at the critical point corresponding to the closure of the gap :
\be
F_K(0) = \frac{N^K}{K!} e^{\frac{NM}{2}}\int \prod_{l=1}^K \frac{db_l}{2\pi}
(1 + b_l^2)^{\frac{M}{2}}\Delta^2(b)e^{-\frac{N}{2}\sum_{l=1}^K b_l^2}
\ee
Note that this expression is exact for finite N.
In the large N limit, we exponentiate the logarithmic term and
expand the exponent about $b_l$ up to order $b_l^4$ term. The critical point,
is precisely the point at which the coeffeicient of the quadratic term
$b_l^2$ vanishes. We then  have
\be
F_K(0) = \frac{N^K}{K!} e^{\frac{NM}{2}}\int \prod \frac{db_l}{2\pi}
e^{-\frac{N}{2}\sum_{l=1}^K b_l^4} \Delta^2(b)
\ee
As in all the cases which appeared in the previous sections, this
integral is expressed by a Hankel determinant, in which the matrix elements
are $\Gamma(\frac{2n-1}{4})$.
The determinant is
\be
I = \det \left|
\matrix{ \Gamma(\frac{1}{4})
&0&\Gamma(\frac{3}{4})&0& \ldots\cr
0&\Gamma(\frac{3}{4})&0&\Gamma(\frac{5}{4})&\ldots\cr
\Gamma(\frac{3}{4})&0&\Gamma(\frac{5}{4})&0&\ldots\cr
\ldots&\ldots&\ldots&\ldots&\ldots}\right|.
\ee

Note that we have considered the case $a_l=\pm 1$ case, but the formulae
are explicit for any
spectrum of the source and they could be easily used to study for instance
 multi-critical situations which were discussed
in \cite{BH3}.

\begin{center}
{\bf Acknowledgement}
\end{center}
  We thank Zeev Rudnick and John Keating for useful discussions.
  This work has been supported by the CREST programme of JST.

\end{document}